\documentclass[prl,twocolumn,nofootinbib,amsfonts]{revtex4}
\usepackage{graphicx,bm}
\newcommand{\be}{\begin{equation}}
\newcommand{\ee}{\end{equation}}

\begin{document}

\title{Relativity principle with a low energy invariant scale}
\author{Jos\'e Manuel Carmona}
\email{jcarmona@unizar.es}
\affiliation{Departamento de F\'{\i}sica Te\'orica,
Universidad de Zaragoza, Zaragoza 50009, Spain}
\author{Jos\'e Luis Cort\'es}
\email{cortes@unizar.es}
\affiliation{Departamento de F\'{\i}sica Te\'orica,
Universidad de Zaragoza, Zaragoza 50009, Spain}

\begin{abstract}
The possibility of a modification of special relativity with an
invariant energy scale playing the role of a minimum energy is
explored. Consistency with the equivalence of different inertial
frames is obtained by an appropriate choice of a non-linear action of
the Lorentz group on momentum space. Limits on the low energy cutoff
from tests of Einstein's theory and possible ways to measure the new
energy scale are discussed. 
\end{abstract}

\maketitle

\section{Introduction}

In recent years a new cosmological paradigm has emerged, in which
our Universe is filled with a ``dark energy'' that causes an
accelerated expansion~\cite{acceleration}. The nature of the dark
energy is not known. 
However, a simple explanation is the existence of a cosmological
constant, or, equivalently, a vacuum energy density, which,
according to experimental data, would be of the order
\be
\rho_V\sim\left(10^{-3}\,\text{eV}\right)^4.
\label{exprho}
\ee
The main advantage of this approach to dark energy is that the
presence of a cosmological constant can be accomodated both in
general relativity and in quantum field theory. However, in classical
general relativity the cosmological constant is a completely free
dimensionful parameter, while from the point of view of particle physics,
it measures the energy density of the vacuum, that is, of the state of
lowest energy, which makes it possible to give estimates for its value.
If we are confident that we can use ordinary quantum field theory
all the way up to the Planck scale, $M_P\sim 10^{19}$\,GeV, then we
expect that
\be
\rho_V^{\text{(Planck)}}\sim\left(10^{28}\,\text{eV}\right)^4.
\ee
There is then a mismatch of 124 orders of magnitude between the expected and
the measured values of the vacuum energy density, or of 31 orders of
magnitude between the Planck mass and the energy scale which defines
the experimental value of $\rho_V$. Even if one thinks that the energy
cutoff of quantum field theory should be changed from the Planck scale to
a lower mass scale, the discrepancy would still be of many orders of magnitude.
In addition, one expects contributions to $\rho_V$ coming from the
different electroweak, chiral or even GUT phase transitions, of the order of
the corresponding energy scales at which these transitions took place
in the early universe. It is very difficult to think why all these
apparently unrelated contributions should almost cancel to produce the
observed value Eq.~(\ref{exprho}).

An alternative to trying to derive the mass scale 
which controls the vacuum energy density in terms of other measurable
parameters is to postulate that it is a fundamental energy scale, just
like $c$ is a fundamental velocity scale. 

Also there is a general agreement that in a quantum
theory of gravity the maximum entropy of any system should be
proportional to the area and not to the volume~\cite{Susskind,Bekenstein}
which implies the breakdown of any effective field theory to describe
systems which exceed a certain critical volume. This observation
leads~\cite{CKN,CC} to endow the effective field theory with an
ojo infrared (IR) cutoff. 
Recently a possible connection between the matter-antimatter asymmetry
of the universe and a violation of CPT parametrized by a low energy
scale has been pointed out~\cite{CCDGM}.

All these arguments lead to consider a new low energy scale as a possible
relic in the microscopic world of the gravitational interactions.

However, an energy scale is not
invariant under Lorentz transformations between inertial observers.
One could ask whether it would be possible to find a modified
relativity principle in which all inertial observers could agree on
the value of the cosmological constant as a fundamental quantity.

In fact attempts along the line to generalize special relativity to
include the invariance of an energy scale have been started to be
explored recently in the framework of the so-called ``Doubly special
relativity''~\cite{DSR1,DSR2}.
Motivated by quantum gravity ideas, the invariant
energy scale has been assumed to be the Planck mass in these attempts.
The fact that the Planck mass is much larger than the energy scales
explored so far in particle physics makes that the new kinematics
is only a very small, high-energy correction to standard relativistic 
kinematics.

In this letter we are questioning however the possibility to find
an ``infrared doubly special relativity'', in the sense that we
explore whether the presence of an energy scale of the order of
$10^{-3}$\,eV, which in principle could be thought to crudely affect
even to nonrelativistic kinematics, could be compatible
with the standard physics we know.

In the attempts to incorporate the Planck energy as a new invariant in
a relativistic theory the main ingredient is to consider a non-linear
realization of Lorentz transformations in momentum space. This
suggests to use the same idea in order to incorporate a new invariant
low energy scale. The simplest way to do that is to take the
non-linear action of Lorentz transformations in DSR2~\cite{DSR2} which
leads to an invariant scale as an ultraviolet cutoff for the energy
and replace everywhere the energy by its inverse. In this way one gets
automatically a non-linear action of Lorentz transformations in
momentum space with an invariant energy scale which now corresponds to
an infrared cutoff as we were looking for. Although this simplest
first choice will turn out to be phenomenologically inconsistent it is
convenient to discuss first this case because most of the results can
be applied to a simple modification which avoids these inconsistencies.

The explicit form of the non-linear Lorentz transformations is defined
by the relations
\be
p_0 = \pi_0 + \lambda \mbox {\hskip 1cm}
{\bm p} = {\bm \pi} \, \left(1+\frac{\lambda}{\pi_0}\right)
\label{nllt}
\ee
where $\pi_0$, ${\bm \pi}$ is a linearly transforming auxiliary
four-momentum and $p_0$, ${\bm p}$ the physical four-momentum. From
these relations one gets
\be
\pi_0 = p_0 - \lambda \mbox {\hskip 1cm}
{\bm \pi} = {\bm p} \, \left(1-\frac{\lambda}{p_0}\right)
\ee
which makes manifest the role of the invariant energy scale $\lambda$ as
a minimum for the physical energy ($p_0>\lambda$). The standard
relativistic invariant mass-shell condition for the auxiliary
four-momentum ($\mu^2=\pi_0^2 - {\bm \pi}^2$) can be reexpressed in
terms of the physical four-momentum as
\be
\mu^2 = \left(1 - \frac{\lambda}{p_0}\right)^2 \left(p_0^2 - {\bm
  p}^2\right) \,.
\label{mash}
\ee

\section{Bounds from QED tests}

Our observations of electromagnetic radiation going down to
frecuencies of the order of Hz or energies of the order of tens of
feV exclude the possibility to have a nonlinear representation of
Lorentz transformations like (\ref{nllt}) for photons unless the
scale $\lambda$ takes an unobservable small value.   

Then in order to proceed one has to consider a generalized relativity
principle which does not affect the kinematics of massless
particles. 
There is a very simple way to modify (\ref{nllt}) in such a way that
only massive particles are sensitive to the low energy scale. One can
consider a nonlinear realization of Lorentz transformations based on  
\be
p_0 = \pi_0 \, \left[1+\frac{\lambda}{\pi_0} f\left(\frac{\pi_0^2-{\bm   
      \pi}^2}{\lambda^2}\right)\right] 
\label{nllt2.1}
\ee
\be
{\bm p} = {\bm \pi} \, \left[1+\frac{\lambda}{\pi_0}
  f\left(\frac{\pi_0^2-{\bm \pi}^2}{\lambda^2}\right)\right]
\label{nllt2.2}
\ee
with a function $f$ such that $f(0)=0$ and $f(\infty)=1$. The first
condition on $f$ leads to linear Lorentz transformations for massless
particles and the second one justifies the approximation in
(\ref{nllt}) for particles whose mass is much larger than the low
energy invariant scale. This will be the case for all the known
elementary particles with the only possible exception of neutrinos for
which it may be neccesary to consider the exact expression
(\ref{nllt2.1}-\ref{nllt2.2}). In fact the net effect of the
modification of the relativity principle is to replace the universal
scale $\lambda$, in the kinematics of any particle with a mass
parameter $\mu$, by 
\be 
\lambda(\mu) = \lambda f\left(\frac{\mu^2}{\lambda^2}\right)
\ee  
which approaches $\lambda$ when $\mu\gg\lambda$.

For all the particles where there is a
good control of the approach to the non-relativistic limit (i.e. for
all the particles except neutrinos) it is clear that the mass scale
$\mu$ should be much greater than the new scale $\lambda$ in order to
reproduce the very well tested relativistic corrections to the
non-relativistic limit. In fact since the corrections due to the new
scale are proportional to the ratio $\lambda/p_0$ the best place to
look for a signal of the new scale is through a small deviation from the
relativistic corrections as predicted by special relativity.
An expansion in powers of the new scale $\lambda$ leads to a
modification of the dispersion relation which to first order in
$\lambda$ is given by
\be
p_0 \simeq \sqrt{{\bm p}^2 + \mu^2} + \frac{\lambda(\mu)}{1+\frac{{\bm
      p}^2}{\mu^2}} \,.
\label{p0}
\ee
In the non-relativistic limit one has
\be
p_0 \simeq m + \frac{{\bm p}^2}{2 m} - \frac{\mu}{m} \,
\frac{{\bm p}^4}{8 m^3}
\ee
where we have introduced a physical mass parameter
\be
m = \mu + \lambda(\mu) \,.
\label{m}
\ee
We see that the first two terms are the same as in the
non-relativistic limit of special relativity, the effect of the new
scale $\lambda$ being reduced to a redefinition of the physical mass
$m$ as a combination of the infrared scale $\lambda$ and the auxiliary
mass parameter $\mu$. But we have a correction in the coefficient of
the next order term by a factor $\mu/m$ and by applying this
kinematical analysis to the electron we would find a slight
modification of the energy levels of the hydrogen atom. Given the
extraordinary agreement between theory and the experimental
measurement of the Lamb shift (one part in $10^5$~\cite{Kino,Wein})
we get a bound on the low energy scale $\lambda\lesssim 5 \, \text{eV}$.

The most stringent bound on the low energy scale that can be obtained
from QED tests comes from the possible modification induced by the
presence of such scale in the determination of the anomalous magnetic
moment of the electron. If one uses the Dirac equation in terms of the
auxiliary variables one can show that at tree level all the effect of
the low energy scale can be reabsorbed in the mass parameter $m$ and
then no anomalous magnetic moment ($a_e$) is generated in this
approximation. A simple estimate of the correction to the usual
calculation imposed by the low energy scale is
\begin{equation}
\label{correccion}
\delta a_e \sim \frac{\alpha}{\pi} \left(\frac{\lambda}{m_e}\right)
\sim 4\times 10^{-9} \frac{\lambda}{(1 \ {\rm eV})}.
\end{equation}
If we ask this correction to be smaller than the uncertainty of the 
theoretical prediction for $a_e$ in QED caused by the uncertainty in the 
determination of $\alpha$~\cite{Kino}, we get a bound for the IR 
scale $\lambda \lesssim 10^{-2}$ eV. 

\section{Bounds from neutrino physics}

Another way to look for bounds on the low energy scale $\lambda$ is to
consider neutrinos. In this case masses can be of the order of the low energy
scale and kinematical corrections can be more important.

One indirect bound on the low energy scale comes
from the contribution of neutrinos to the energy density of the
Universe~\cite{PDG} which puts a bound on the physical mass parameter
$m$ of the neutrino. If one assumes three degenerate Dirac neutrinos
one gets $m < 4 \,\text{eV}$. Since the low energy scale $\lambda(\mu)$ is
bounded by $m$ (as a consequence of the definition of the physical
mass parameter in (\ref{m}) and the condition $\mu>0$) then the upper
bound on $m$ is directly an upper bound on $\lambda(\mu)$.

Another indirect bound on the low energy scale comes from experiments
trying to measure or put limits on neutrino masses from the tritium
beta-decay spectrum which at present~\cite{PDG} are at the level of
the eV. The bound on the low energy scale will be at best of the same
order of magnitude.

Then the indirect bounds on the low energy scale from bounds on
neutrino masses are less stringent than the bound obtained from QED
tests.  

If one tries to get direct bounds from the observations of neutrinos 
there is an additional difficulty to identify a signal of the low energy
scale. Only neutrinos whose energy is much larger than its mass can be
detected and once more the kinematical corrections due to the low
energy scale are suppressed. We need in this case the
ultra-relativistic limit of (\ref{p0}): 
\be
p_0 \simeq |{\bm p}| + \frac{\mu^2}{2 |{\bm p}|} +
\frac{\lambda(\mu) \mu^2}{{\bm p}^2}\,.
\label{url}
\ee
The phase of the oscillating amplitude between two neutrino states with
mass parameters $\mu_1$ and $\mu_2$ will be
\be
\theta_{12} = L (E_1 - E_2) \approx L \frac{\mu_1^2 -\mu_2^2}{2|{\bm p}|} + L 
\frac{\mu_1^2 \lambda(\mu_1) -\mu_2^2 \lambda(\mu_2)}{{\bm p}^2}
\,.
\ee
If one has an experiment with the ratio $|{\bm p}|/L$ of the order of
$\mu_1^2 - \mu_2^2$ one will observe the oscillation between the two types of
neutrinos as expected in special relativity and the correction due to
the infrared scale $\lambda$ will be unobservable. In order to have a
signal of the infrared scale in neutrino oscillations one requires an
experiment with ${\bm p}^2/L$ of the order of $\mu_1^2 \lambda(\mu_1)
-\mu_2^2 \lambda(\mu_2)$. The ratio ${\bm p}^2/L$ can be written as
\be
\frac{{\bm p}^2}{L} \approx 2 \times 10^{-6}\;{{\rm eV}}^3
\left(\frac{|{\bm p}|}{{\rm MeV}}\right)^2 \left(\frac{10^8 \,{\rm
    Km}}{L}\right) \,. 
\ee
Then one sees that solar neutrino experiments are the appropriate
place to look for a signal of the low energy scale or for a relevant
bound on it.

\section{Summary and discussion}

If one wants to go beyond this simple kinematical analysis of possible
signatures of a low energy cutoff one should have to consider a
dynamical theory where the non-linear Lorentz transformations, as
defined by (\ref{nllt}), were realized on the space of states. This is
an open problem which has been the subject of recent work~\cite{We,Ame}
in connection with a modified relativity principle with a high energy
invariant. It is not possible at present to do a dynamical analysis,
it is not even clear whether there can be an obstruction to construct
a dynamical theory with a modified relativity principle compatible
with an invariant energy scale.

Another limitation of the present work is that all the analysis has
been concentrated on the non-linear realization of Lorentz
transformations based on the introduction of the physical momentum
variables as in (\ref{nllt}). This is not the only possible way to
introduce a low energy scale. It corresponds to a translation to low
energies of a similar construction at high energies~\cite{DSR2} but
there are many other possibilities. We expect that the main conclusions
are not a peculiarity of the choice of the non-linear transformations used
and that they will apply to a more general case.

To summarize we have found that it is possible to introduce a low
energy invariant scale in a relativistic theory and that if this
new scale is of the order of $10^{-3}\text{eV}$ it can be made compatible
with all the tests of special relativity and at the same time one is
close to detect its signals in different experiments. If one assumes a
mechanism of cancellation for the contributions to the vacuum
expectation value of the energy momentum tensor then on dimensional
grounds one would expect  
\be
<T_{\mu\nu}> \sim \lambda^4 \left(c_1 \delta_{\mu}^0
\delta_{\nu}^0 + c_2 \sum_i \delta_{\mu}^i \delta_{\nu}^i\right)
\ee
with $c_1$, $c_2$ dimensionless coefficients depending on the details
of the theory incorporating the new low energy scale. Then the present
acceleration of the expansion of the Universe could be a signal of a
new low energy scale compatible with a modified relativity principle.

We are grateful to Jorge Gamboa for a collaboration in the first steps
of this work and to Victor Laliena and Stefano Liberati for
discussions. This work has been supported by project FPA2003-02948 , by
Fondecyt, an by an INFN-CICyT collaboration grant.


\end{document}